\documentclass[twocolumn,secnumarabic,amssymb,nobibnotes,aps,superscriptaddress,pra]{revtex4-1}
\pdfoutput=1 

\usepackage[dvips]{graphicx}
\usepackage{bm}% bold math
\usepackage{latexsym} 
\usepackage{amsmath}
\usepackage{amssymb}
\usepackage{color}

\usepackage{eucal}

\newcommand{\ui}{{\rm i}}
\newcommand{\eps}{{\varepsilon}}

\newcommand{\bmA}{{\bm A}}

\newcommand{\bmr}{{\bm r}}
\newcommand{\bmp}{{\bm p}}

\newcommand{\bmv}{{\bm v}}

\newcommand{\bms}{{\bm s}}

\newcommand{\bmQ}{{\bm Q}}

\newcommand{\bra}{\langle}
\newcommand{\ket}{\rangle}
\newcommand{\kB}{k_{\rm B}}

\makeatletter
\renewcommand*{\p@subsection}{}
\renewcommand*{\p@subsubsection}{}
\makeatother

\begin{document}

\title{
  Spin Hall effect generated by fluctuating vortices in type-II superconductors 
}

\author{Takuya Taira} 
\affiliation{Department of Physics, Okayama University, Okayama 700-8530, Japan} 
\author{Yusuke Kato} 
\affiliation{Department of Basic Science, University of Tokyo, Meguro, Tokyo 153-8902, Japan}
\author{Masanori Ichioka}
\affiliation{Research Institute for Interdisciplinary Science, Okayama University, Okayama 700-8530, Japan}
\affiliation{Department of Physics, Okayama University, Okayama 700-8530, Japan}
\author{Hiroto Adachi}
\affiliation{Research Institute for Interdisciplinary Science, Okayama University, Okayama 700-8530, Japan}
\affiliation{Department of Physics, Okayama University, Okayama 700-8530, Japan}
\date{\today}

\begin{abstract}
  We theoretically investigate the vortex spin Hall effect, i.e., a novel spin Hall effect driven by the motion of superconducting vortices, by focusing on the role of superconducting fluctuations. Within the BCS-Gor'kov microscopic approach combined with the Kubo formula, we find a strong similarity between the vortex spin Hall effect and the vortex Nernst/Ettingshausen effect. Calculated temperature dependence of the voltage signal due to the inverse vortex spin Hall effect exhibits a strong enhancement by vortex fluctuations. This result not only provides a possible explanation for a prominent peak found in the spin Seebeck effect in a NbN/Y$_3$Fe$_5$O$_{12}$ system, but also leads to a proposal of new experiments using other superconductors with strong fluctuations, such as cuprate or iron-based superconductors. 
\end{abstract} 


%display desired

\maketitle

%%%%%%%%%%%%%%%%%%%%%%%%%%%%%%%%%%%%
\section{Introduction \label{Sec:I}}
%%%%%%%%%%%%%%%%%%%%%%%%%%%%%%%%%%%%
Since Aronov discussed spin injection into superconductors~\cite{Aronov77}, spin transport in superconductors has been a subject of intense research both theoretically~\cite{Yafet83,Kivelson90,Zhao95,Yamashita02,Takahashi02,Morten08,Takahashi08,Inoue17,Bergeret18,Montiel18,Taira18,Kato19} and experimentally~\cite{Yang10,Hubler12,Quay13,Wakamura14,Ohnishi14,Yao18,Umeda18,Jeon18}, leading to the emergence of superconducting spintronics~\cite{Linder15,Eschrig15,Ohnishi20}. Although the order parameters in a spin-triplet superfluid are spin polarized and therefore the condensate itself can carry spins~\cite{Serene83}, most of the superconductors have spin-singlet pairing with zero spin polarization. So, at first glance, a spin current in a spin-singlet superconductor is inevitably carried by the excitations of Bogoliubov quasiparticles, and the dynamics of the superconducting order parameter does not seem to play a major role. For this reason, pursuing a spin current that is carried by the condensate in an abundant spin-singlet superconductors is one of the most challenging issues in superconducting spintronics. 

\begin{figure}[t] 
  \begin{center}
    \includegraphics[width=6cm]{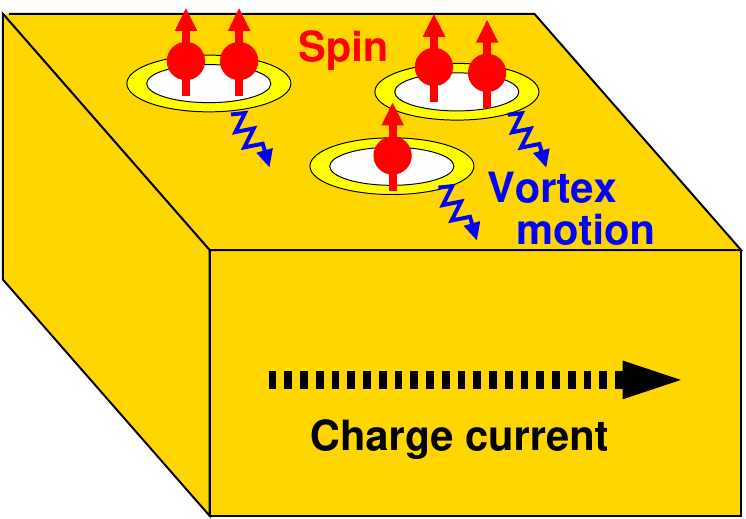}
  \end{center}
  \caption{Schematic drawing of the vortex spin Hall effect discussed in this paper. A superconducting vortex hosts spin accumulation in the core, which moves transverse to the charge current following the Josephson equation~\cite{Josephson62,Anderson66}. Compare the drawing with Fig. 1 of Ref.~\cite{Wang06}. } 
  \label{fig:schematic01}
\end{figure}

As stated above, a spin-singlet Cooper pair does not host a nonzero spin polarization, and it is not so easy to think of a spin current carried by the superconducting order parameter. The situation is similar to that of a heat current in a superconductor. It is known that in the Meissner state, no entropy is transferred by the supercurrent~\cite{Landau-fluid}, and the heat is carried only by Bogoliubov quasiparticles. However, in the vortex state of a type-II superconductor, the situation is drastically changed. In this case, the vortices can host a nonzero entropy, and the vortex motion can contribute to the heat transport. Since the vortices move approximately transverse to the charge current direction following the Josephson equation~\cite{Josephson62,Anderson66}, this vortex motion produces a transverse heat flow. Therefore, the vortex motion is the source of a large Nernst/Ettingshausen effects in the superconducting vortex state~\cite{Wang06,Pourret06}. Because the superconducting vortex can host a spin polarization as well by a static bias due to the paramagnetic effect~\cite{Maki66,Caroli66,Adachi05,Ichioka07} or by a dynamic stimulus due to the spin pumping~\cite{Inoue17}, it is natural to think about replacing the vortex entropy by a vortex spin polarization in order to discuss a spin current in the vortex state (Fig.~\ref{fig:schematic01}).  

Recently, along the lines of the idea mentioned above, the spin polarization moving with superconducting vortices is discussed by focusing on the flux flow regime~\cite{Vargunin19}. Although the idea having recourse to the flux flow picture is conceptually transparent, when comparing the theoretical result with experiments, one faces the following difficulty~\cite{Tinkham-textbook}. That is, the flux flow theory does not explain {\it temperature} dependence of the resistivity in a type-II superconductor under magnetic fields. Instead, it is the fluctuation theory~\cite{Skocpol75,Larkin-textbook} that can explain a broadening of the resistive transition, as proven not only in conventional superconductors~\cite{Masker69,Hassing73} but also in cuprate superconductors~\cite{Ikeda89,Ullah91}. Therefore, instead of referring to this phenomena as flux flow spin Hall effect, it is more adequate to call it vortex spin Hall effect. Then, it is highly demanded to analyze the vortex spin Hall effect on the basis of the fluctuation theory. 

In this paper we theoretically explore the vortex spin Hall effect, with special focus on the influence of superconducting fluctuations. Using the Kubo formula, we microscopically calculate the spin Hall conductivity in a vortex fluctuation region. A similar physical situation has been dealt with in Ref.~\cite{Kim18} but the discussion is highly phenomenological, which is contrasted to our microscopic approach. Specifically, by microscopically calculating the spin current vertex, we reveal that there is a similarity between this vortex spin Hall effect and the well-known vortex Nernst/Ettingshausen effect~\cite{Wang06,Pourret06}. The origin of this similarity is explained by an intuitive picture that the spin polarization is transported by vortices in the former, whereas the spin polarization is replaced by entropy in the latter (Fig.~\ref{fig:schematic01}). Furthermore, we calculate temperature dependence of the voltage signal generated by the inverse vortex spin Hall effect. This allows us to compare the theory with experiments. In doing so, we pay attention to the fact that the validity of the fluctuation theory grows substantially in the vortex state as the fluctuations acquire one-dimensional character, such that we need to include non-Gaussian fluctuations~\cite{Bray74,Thouless75}. 

This work is motivated by a recent longitudinal spin Seebeck experiment in a bilayer of a superconductor NbN and a ferrimagnetic insulator Y$_3$Fe$_5$O$_{12}$ (YIG)~\cite{Umeda18}. There, a pronounced peak in the inverse spin Hall signal was observed at the superconducting transition, and the result was interpreted in terms of the coherence peak effect predicted for the spin pumping into superconductors~\cite{Inoue17}. However, the interpretation in terms of the coherence peak effect was subsequently questioned~\cite{Kato19}, since the coherence peak effect exists only in an extremely low energy scale much less than the superconducting gap, whereas the spin Seebeck effect consists of spectrally broad components extended up to an energy scale much greater than the superconducting gap. This means that the coherence peak effect is expected to be drowned out completely in the case of the spin Seebeck effect. The present work provides us with an alternative interpretation of the experiment~\cite{Umeda18} in terms of the vortex spin Hall effect. 

The rest of this paper is organized as follows. In Sec.~\ref{Sec:II}, we explain the formalism we use and present a microscopic calculation of the spin current vertex by comparing it with the charge and heat current vertices. In Sec.~\ref{Sec:III}, we present our microscopic calculation of the vortex spin Hall conductivity. In Sec.~\ref{Sec:IV}, taking account of the vortex fluctuation effects which are indispensable to explain temperature dependence of the resistive transition, we calculate temperature dependence of the inverse spin Hall voltage. Finally in Sec.~\ref{Sec:V}, we summarize our result. 

We use the unit $\hbar = \kB = c= 1$ throughout this paper.

%%%%%%%%%%%%%%%%%%%%%%%%%%%%%%%%%%%%% 
%%%%%%%%%%%%%%%%%%%%%%%%%%%%%%%%%%%%%

%%%%%%%%%%%%%%%%%%%%%%%%%%%%%%%%%%%%
\section{Formulation and current vertices \label{Sec:II}}
%%%%%%%%%%%%%%%%%%%%%%%%%%%%%%%%%%%%eg

We begin this section with the following Hamiltonian for an $s$-wave superconductor~\cite{Maki68,Ebisawa71,Werthamer-review}: 
%%%
\begin{eqnarray}
  {\cal H} &=& {\cal H}_{0} - |g| \int d^3 r \Psi^\dag (\bmr)  \Psi (\bmr)
  \label{eq:H_BCS01}
\end{eqnarray}
%%%
where $|g|$ is the BCS attractive interaction parameter, and $\Psi(\bmr) = \psi_- (\bmr) \psi_+ (\bmr)$ is the pair field with electron field operator $\psi_\sigma (\bmr)$ for spin projection $\sigma=\pm$. In Eq.~(\ref{eq:H_BCS01}), 
%%%
\begin{eqnarray}
  {\cal H}_{0} &=& \sum_{\sigma} \int d^3 r \psi^\dag_\sigma (\bmr) \left[ 
  \frac{(-\ui {\bm \nabla} + |e| {\bm A} )^2}{2m}  
  + U(\bmr) 
  \right]
  \psi_\sigma (\bmr) \nonumber \\
  \label{eq:H0_01}  
\end{eqnarray}
%%%
is the single particle Hamiltonian, where ${\bm A}$ is the vector potential, $m$ and $|e|$ are the mass and absolute value of charge of an electron. The potential $U(\bmr)= \sum_a U_{\rm imp} (\bmr- \bmr_a)$ is due to impurities at fixed positions $\bmr_a$, and after the impurity average denoted by an overline, it has zero mean and variance $\overline{U(\bmr)U(\bmr')}= (2 \pi N(0) \tau)^{-1} \delta (\bmr- \bmr')$, where $N(0)$ and $\tau$ are the density of states per spin and electron lifetime, respectively. Note that the above model does not include any spin-orbit interactions.

For the calculation of the spin Hall conductivity in a fluctuating vortex state, we use the Kubo formula~\cite{Crepieux01,Tse06,Okamoto19}: 
%%%
\begin{equation}
  \sigma^{\rm (SH)}_{xy}
  =
  \frac{1}{\ui \Omega} \left. \left( K^{{\rm (SH)} R}_{xy} (\Omega) - K^{{\rm (SH)} R}_{xy} (0)\right) \right|_{\Omega \to 0},
  \label{eq:Kubo_SH01}
\end{equation}
%%%
where the retarded correlation function $K^{{\rm (SH)} R}_{xy} (\Omega)$ is obtained from the Matsubara correlation function, 
%%%
\begin{eqnarray}
K_{xy}^{{\rm (SH)}}(\ui \Omega_{m}) &=& 
  \frac{1}{V}  \int d^3 r \int d^3 r'  
  \int_0^{\beta} d u e^{\ui \Omega_{m} u} \nonumber \\
  && \times \bra T_u \; {\mathcal J}^{(s)}_x (\bmr, u) {\mathcal J}^{(c)}_y (\bmr', 0) \ket,
  \label{eq:Kxy_func01}
\end{eqnarray}
%%%
through the analytic continuation $K^{{\rm (SH)}R}_{xy} (\Omega) = K^{\rm (SH)}_{xy} (\ui \Omega_{m} \to \Omega+ i 0_+)$. In Eq.~(\ref{eq:Kxy_func01}), $V$ is the volume of the system, $\beta= 1/T$ is the inverse temperature, $u$ is the imaginary time, and $T_u$ is the time ordering operator. The charge and spin current operators are, respectively, given by 
%%%
\begin{eqnarray}
  {\bm {\mathcal J}}^{(c)} &=&
    \frac{-|e|}{2m} \sum_{\sigma} \psi^\dag_\sigma \left( -\ui {\bm \nabla}+ |e| \bmA \right) \psi_\sigma + H. c., \\
  {\bm {\mathcal J}}^{(s)} &=&
  \frac{-|e|}{2m} \sum_{\sigma} \sigma \psi^\dag_\sigma \left( -\ui {\bm \nabla}+ |e| \bmA \right) \psi_\sigma + H. c., 
\end{eqnarray}
%%%
where $H.c.$ denotes Hermitian conjugate. Note that, following convention, we define the spin current so as to have the same dimension as the charge current. 

We assume the dirty limit $1/(2 \pi T_{\rm c0} \tau) \gg 1$, where $T_{\rm c0}$ is the superconducting transition temperature under zero magnetic field. In our diagrammatic calculation, we use the familiar quasiclassical approximation~\cite{Werthamer-review} for the impurity-averaged electron Green's function for spin projection $\sigma$ under a magnetic field ${\bm H}= H \hat{\bf z}$: 
%%%
\begin{equation}
  G^H_{\eps_n, \sigma}(\bmr, \bmr')
  = G_{\eps_n, \sigma}(\bmr-\bmr') \exp{\Big( \ui |e| \int_{\bmr}^{\bmr'} \bmA \cdot d \bms \Big)}, 
\end{equation}
%%%
where $G_{\eps_n, \sigma}(\bmr-\bmr')$ is the impurity-averaged Green's function under zero magnetic field. This quantity is given by the Fourier transform of 
%%%
\begin{eqnarray}
  G_{\bmp, \sigma}(\eps_n)
  &=& \frac{1}{\ui \widetilde{\eps}_n - \epsilon_\bmp+ \mu_\sigma} 
  = \frac{1}{\ui \widetilde{\eps}_n - \xi_\bmp+ \sigma {\mu}_s/2},
  \label{eq:Green01}
\end{eqnarray}
%%%
where $\widetilde{\eps}_n= \eps_n + {\rm sgn}(\eps_n)/2 \tau$, and $\eps_n= 2 \pi T (n+1/2)$ is a fermionic Matsubara frequency. In Eq.~(\ref{eq:Green01}), $\epsilon_\bmp= \bmp^2/(2m)$, $\mu_\sigma$ is the spin-dependent chemical potential, $\xi_\bmp= \epsilon_\bmp- \mu_0$, $\mu_0 = (\mu_+ + \mu_-)/2$, and ${\mu}_s = (\mu_+ - \mu_-)$ is the spin accumulation. Note that we assume the presence of a nonzero spin accumulation ${\mu}_s$ by a stimulus of the spin pumping or the spin Seebeck effect. In the following, we consider a particle-hole symmetric case, and we investigate terms up to the linear order with respect to the spin accumulation ${\mu}_s$.

%%%%%%%%%%%%%%%%%%%%%%%%%%%%%%%%%%%%% 
\begin{figure}[t] 
  \begin{center}
        \includegraphics[width=7.0cm]{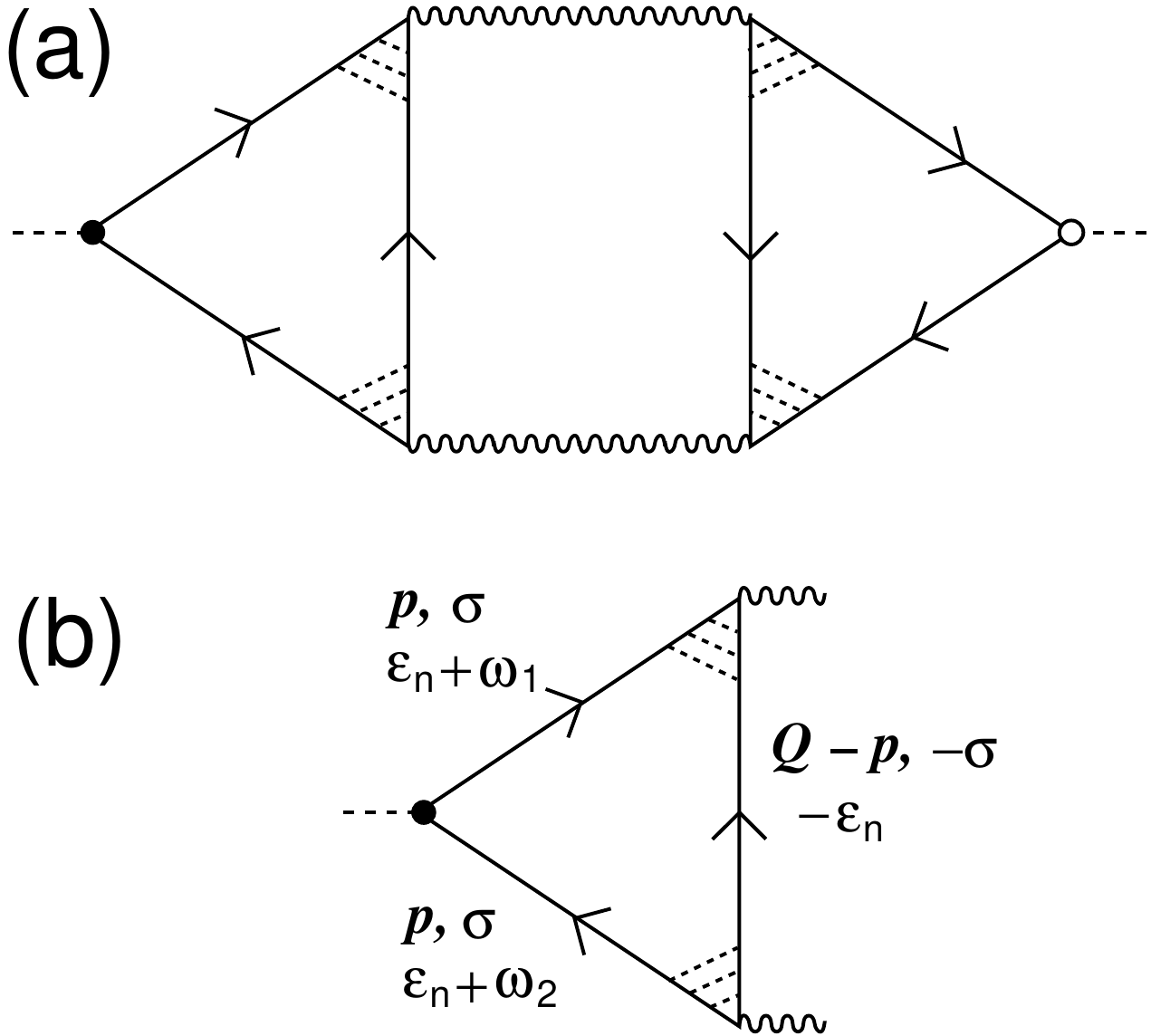}
  \end{center}
  \caption{(a) Diagrammatic representation of the process considered in this work. A wavy line represents the fluctuation propagator, a solid line with an arrow is the electron Green's function. Dashed lines connected to a filled circle and an open circle are the fermionic spin current and charge current vertices, respectively. A dashed ladder denotes a Cooperon vertex [Eq.~(\ref{eq:Cooperon01})]. (b) Diagram representing the bosonic spin current vertex acting on the pair field. In the case of the charge current vertex, the filled circle is replaced by an open circle.} 
  \label{fig:triangle01}
\end{figure}
%%%%%%%%%%%%%%%%%%%%%%%%%%%%%%%%%%%%%

For the calculation of the vortex spin Hall effect in the fluctuating vortex state, we focus on the process shown in Fig.~\ref{fig:triangle01}(a). If a Gaussian approximation valid above the mean-field transition temperature is used for the fluctuation propagator, the process coincides with the Aslamazov-Larkin contribution~\cite{AL68}. If, on the other hand, a one-loop self-energy correction to the fluctuation propagator is taken into account in a self-consistent way (see Fig.~2(a) of Ref.~\cite{Ikeda89}), the process gives us the so-called self-consistent Hartree approximation~\cite{Ullah91} that is valid both above and below the mean-field transition temperature in the vortex state. In this work we adopt the latter approximation, i.e., the self-consistent Hartree approximation~\cite{Ullah91}. As we show below, there is a similarity between the vortex spin Hall effect under discussion and the well-known vortex Nernst/Ettingshausen effect~\cite{Maki71,Ussishkin02,Serbyn09,Michaeli09}. For the latter effect, the process shown in Fig~\ref{fig:triangle01}(a), which coincides with the Aslamazov-Larkin process within the Gaussian approximation, is known to be dominant among other contributions~\cite{Ussishkin03}. Therefore, considering the similarity between the vortex spin Hall effect and vortex Nernst/Ettingshausen effect, we focus on this process [Fig.~\ref{fig:triangle01}(a)]. The Maki-Thompson process and the density of states process are disregarded.

  In the remaining part of this section, we calculate the bosonic current vertex acting on the pair field, which is constructed from the fermionic triangle shown in Fig.~\ref{fig:triangle01}(b)~\cite{Usadel69,Aronov95,Ussishkin03}. Note that in our calculation, we use the following identity~\cite{Werthamer-review}: 
% It is known that, within a Gaussian approximation,} the Aslamazov-Larkin process corresponds to the contribution from fluctuations of the superconducting order parameter, whose dynamics is described by the time-dependent Ginzburg-Landau (TDGL) equation~\cite{Maki71,Ussishkin02}. Also known is that the fermionic triangle constitutes the bosonic current vertex acting on the pair field~\cite{Usadel69,Aronov95,Ussishkin03}. Note that in our calculation, we use the following identity~\cite{Werthamer-review}: 
%%%
\begin{eqnarray}
  \exp{\Big( 2 \ui |e| \int_{\bmr'}^\bmr \bmA \cdot d \bms \Big)} \Psi(\bmr')
  = e^{ -\ui (\bmr- \bmr') \cdot \bmQ(\bmr) } \Psi(\bmr),
  \label{eq:gauge-grad01}
\end{eqnarray}
%%%
where $\Psi$ is the pair field, and $\bmQ(\bmr)= -\ui{\bm \nabla} + 2|e|\bmA$ is the gauge-invariant gradient. 

With these in mind, let us first review the calculation of the bosonic charge current vertex shown in Fig.~\ref{fig:triangle01}(b)~\cite{Aronov95}. Starting from the position representation of the fermionic charge current vertex~\cite{Caroli67,Usadel69} and then use Eq.~(\ref{eq:gauge-grad01}), the fermion triangle in the Matsubara representation can be calculated as 
%%%
\begin{eqnarray} 
  {\bf J}^{(c)}  &=& {|e|} 
  T \sum_{\eps_n} \sum_\sigma  \int_\bmp \bmv_\bmp  \Big( G_{\bmQ- \bmp, -\sigma}(-\eps_n) \nonumber \\  
   &\times& G_{\bmp, \sigma}(\eps_n+ \omega_1 ) G_{\bmp, \sigma}(\eps_n+ \omega_2) \Big) \nonumber \\  
  &\times& C_{\bmQ, \sigma}(\eps_n + \omega_1, -\eps_n)
  C_{\bmQ, \sigma}(\eps_n+ \omega_2, -\eps_n), \nonumber \\ 
  \label{eq:Jc01}
\end{eqnarray}
%%%
where $\bmv_\bmp= \bmp/m$, and we introduced the shorthand notation $\int_\bmp= \int d^3 p/(2 \pi)^3$. In this equation,   
%%%
\begin{eqnarray}
  C_{\bmQ,\sigma} (\eps_n+ \omega_\nu, -\eps_n)
  =
    \frac{ \tau^{-1} \Theta \big(\eps_n(\eps_n+ \omega_\nu) \big)}
         {d_\bmQ(2 \eps_n+ \omega_\nu) - \ui \sigma {\mu}_s {\rm sgn}(\eps_n)  }
         \hspace{0.5cm}
         \label{eq:Cooperon01}
\end{eqnarray}
%%%
is the Cooperon vertex, where $\Theta(x)$ is the step function, $d_\bmQ(2 \eps_n+ \omega_\nu)= |2 \eps_n+ \omega_\nu|+ DQ^2$ with $D= v_F^2 \tau/3$ being the diffusion coefficient.

We first notice that dependence on the small bosonic Matsubara frequency $\omega_{1,2} (\ll \eps_n)$ is unimportant in the present case, such that we can safely set $\omega_{1,2} \to 0$. Then, the momentum integral for the product of three Green's functions yields $- 4 \pi N(0) \tau^2 D \bmQ$, and the charge current vertex up to the linear order in $\bmQ$ can be expressed as~\cite{Aronov95,Ussishkin03}
%%%
\begin{equation}
  {\bf J}^{(c)} = t^{(c)} N(0) \bmQ, 
\end{equation}
%%%
where the amplitude $t^{(c)}$ is given by 
%%%
\begin{eqnarray}
  t^{(c)} &=&
  -16 \pi |e| D T \sum_{n=0}^\infty \left(\frac{1}{2 \eps_n} \right)^2 \nonumber \\
  &=&  -4 |e| \xi_0^2, 
\end{eqnarray}
%%%
where $\xi_0^2= \pi D/(8 T_{\rm c0})$ is the coherence length, and in moving to the last line we used $\varPsi^{(1)}(1/2)= \pi^2/2$ for the tri-gamma function $\varPsi^{(1)}(z)$. 

Let us now calculate the bosonic spin current vertex acting on the pair field. The expression for the spin current vertex in the Matsubara representation is given by 
%%%
\begin{eqnarray} 
  {\bf J}^{(s)} &=& {|e|}
  T \sum_{\eps_n} \sum_\sigma  \sigma \int_\bmp \bmv_\bmp  \Big( G_{\bmQ- \bmp, -\sigma}(-\eps_n) \nonumber \\  
   &\times& G_{\bmp, \sigma}(\eps_n+ \omega_1 ) G_{\bmp, \sigma}(\eps_n+ \omega_2) \Big) \nonumber \\  
  &\times& C_{\bmQ, \sigma}(\eps_n + \omega_1, -\eps_n) C_{\bmQ, \sigma}(\eps_n+ \omega_2, -\eps_n), 
  \label{eq:Js01}
\end{eqnarray}
%%%
where the essential change from Eq.~(\ref{eq:Jc01}) is the presence of an additional $\sigma$ in the summand. Although the expression does not seem to differ so much from that of the charge current vertex ${\bf J}^{(c)}$ [Eq.~(\ref{eq:Jc01})], the crucial difference is that now the frequency dependence is of great importance. This is because if we only consider a spin current vertex independent of $\omega_{1,2}$, then the result becomes quite similar to that of the vortex Hall effect~\cite{Fukuyama71}, which is known to vanish in the present particle-hole symmetric case~\cite{Aronov95}. Therefore, we need to evaluate the spin current vertex that is proportional to the frequency $\omega_{1}$ and $\omega_2$. 

Then, expanding the spin current vertex up to the linear order in $\bmQ$, ${\bf J}^{(s)}$ can be expressed as 
%%%
\begin{equation}
    {\bf J}^{(s)}  =  t^{(s)}(\ui \omega_1, \ui \omega_2) N(0) \bmQ, 
\end{equation}
%%%
where $t^{(s)}(\ui \omega_1, \ui \omega_2)$ is the amplitude in the Matsubara space. Now we consider how the analytic continuation with respect to $\ui \omega_1$ and $\ui \omega_2$ is performed. As detailed in Appendix~\ref{Sec:App01}, the way of analytic continuation needed here is that $\ui \omega_1$ ($\ui \omega_2$) is from the upper (lower) half plane to the real axis, i.e., $\ui \omega_{1,2} \to \omega \pm \ui 0_+ $, just like the case of the Coulomb drag~\cite{Kamenev95} (see Fig. 3(b) of Ref.~\cite{Kamenev95}). This means that $\omega_1$ and $\omega_2$ have opposite signs in the Matsubara space.

%%%%%%%%%%%%%%%%%%%%%%%%%%%%%%%%%%%%% 
\begin{figure}[t] 
  \begin{center}
    \includegraphics[width=9cm]{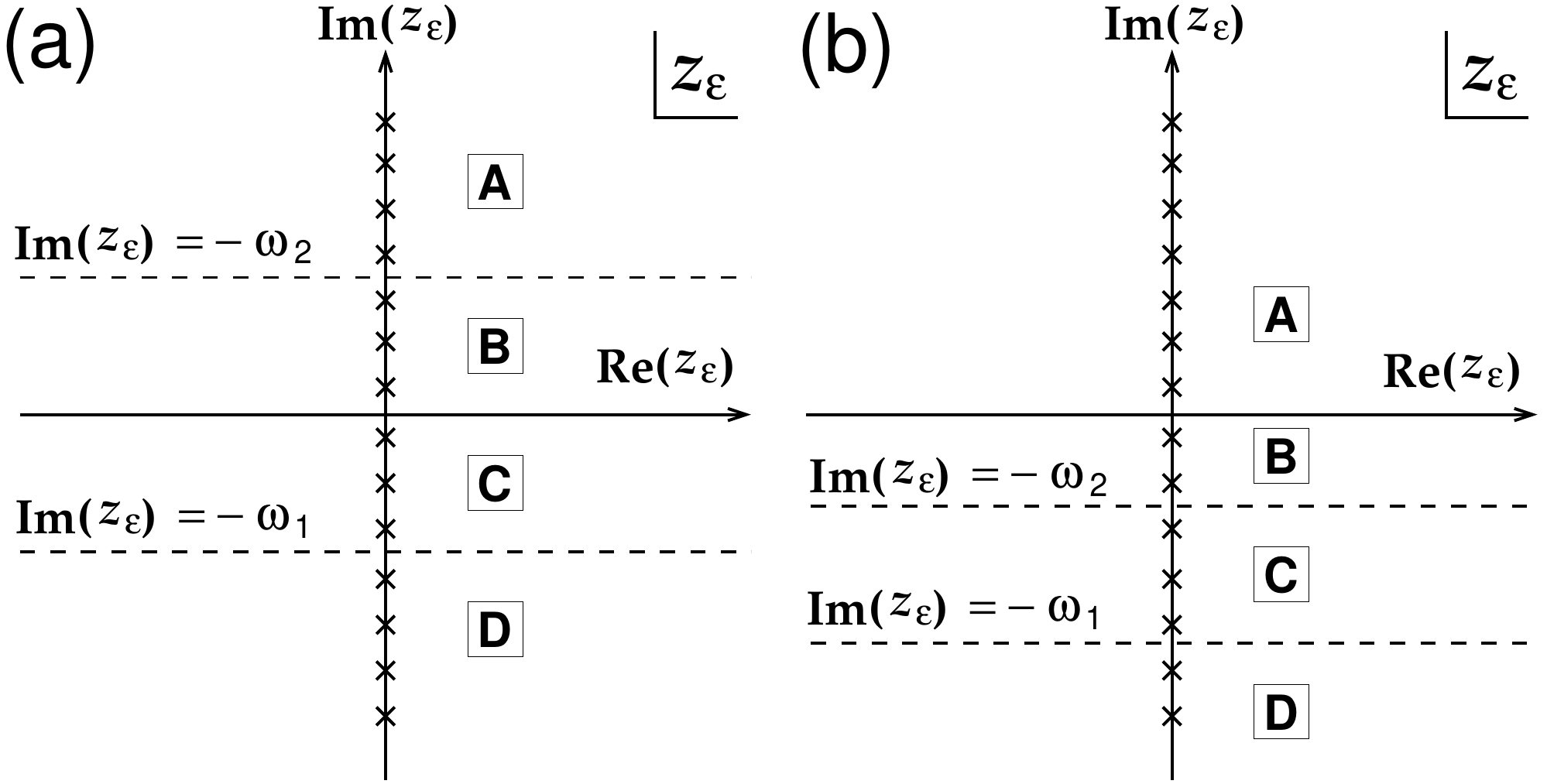}
  \end{center}
  \caption{Four regions \fbox{A}, \fbox{B}, \fbox{C}, and \fbox{D} for Matsubara summation in the complex plane. (a) When $\omega_1 >0$, $\omega_2 < 0$, and (b) $\omega_1>0$, $\omega_2>0$. }
  \label{fig:zeps01}
\end{figure}
%%%%%%%%%%%%%%%%%%%%%%%%%%%%%%%%%%%%%

In performing the $\eps_n$-summation in Eq.~(\ref{eq:Js01}), we have four regions separated by three branch cuts as shown in Fig.~\ref{fig:zeps01}. Because the Cooperon enhancement is the most effective in regions $\fbox{A}$ and $\fbox{D}$, it is enough to evaluate the contribution from these two regions. Carrying out the Matsubara summation over $\eps_n$ and then performing the relevant analytic continuation, whose procedure is detailed in Appendix~\ref{Sec:App01}, the spin current vertex is obtained as 
%%%
\begin{equation}
  {\bf J}^{(s)}(\omega)  =  t^{(s)RA} (\omega,\omega) N(0) \bmQ. 
  \label{eq:Js_02}
\end{equation}
%%%
Here, the amplitude $t^{(s)RA}(\omega, \omega) = t'^{(s)} \omega$ is linear in $\omega$ with the slope given by 
%%%
\begin{eqnarray}
  {t}'^{(s)} &=& {-4 |e|} \xi_0^2 \frac{\mu_s }{8 T^2}, 
\end{eqnarray}
%%%
where the superscript $RA$ in $t^{(s)RA}(\omega, \omega)$ denotes how the analytic continuation with respect to $\ui \omega_1$ and $\ui \omega_2$ is performed. 

It is important to note that we can relate this expression for ${\bf J}^{(s)}$ to the charge current vertex ${\bf J}^{(c)}$ by 
%%%
\begin{eqnarray}   
  {\bf J}^{(s)} (\omega) &=& \frac{\mu_s}{8 T^2} \omega {\bf J}^{(c)} , 
  \label{eq:Js_03}
\end{eqnarray}
%%%
which should then be compared to the expression for the heat current vertex ${\bf J}^{(h)}$~\cite{Ussishkin03},
%%%
\begin{eqnarray}   
  {\bf J}^{(h)} (\omega) &=& \frac{1}{-2|e|} \omega {\bf J}^{(c)}.  
  \label{eq:Jh01}
\end{eqnarray}
%%%
Note also that the spin current vertex in Eq.~(\ref{eq:Js_02}) vanishes when $\omega_1$ and $\omega_2$ have the same sign. Therefore, it survives only in region \fbox{F} in Fig.~\ref{fig:zplane01b} (see the next section). Keeping the latter constraint in mind, we have an important relation between the spin current vertex and the heat current vertex, 
%%%
\begin{eqnarray}   
  {\bf J}^{(s)} (\omega) &=& \frac{-|e|\mu_s}{4T^2} {\bf J}^{(h)}. 
  \label{eq:JsJh01}
\end{eqnarray}
%%%
This result suggests that the vortex spin Hall effect (given by the {\it spin current}-charge current correlation function) has a similarity to the vortex Nernst/Ettingshausen effect (given by the {\it heat current}-charge current correlation function). Besides, we see that the spin current vertex is nonvanishing only in the presence of the spin accumulation $\mu_s$. Note that, in this work the presence of a nonzero ${\mu}_s$ due to the spin pumping or the spin Seebeck effect is assumed, such that $\mu_s$ is {\it not} the quantity to be determined self-consistently. 

%%%%%%%%%%%%%%%%%%%%%%%%%%%%%%%%%%%%%%%%%%%%%%%%%%%%%%%%%%%%%
\section{Vortex spin Hall conductivity\label{Sec:III}}
%%%%%%%%%%%%%%%%%%%%%%%%%%%%%%%%%%%%%%%%%%%%%%%%%%%%%%%%%%%%%
Since the expressions for the spin and charge current vertices are obtained in the previous section, we are now in a position to calculate the vortex spin Hall conductivity. Let us begin with the expression for the total spin Hall conductivity: 
%%%
\begin{equation}
  \sigma^{\rm (SH)}_{xy} = \sigma^{\rm (SH)}_{xy, N}+ \delta \sigma^{\rm (SH)}_{xy},
  \label{eq:sigmaSH01}
\end{equation}
%%%
where $\sigma^{\rm (SH)}_{xy, N}$ is the normal state contribution, whereas $\delta \sigma^{\rm (SH)}_{xy}$ is the fluctuation contribution. Correspondingly, we have two correlation functions $K^{\rm (SH)}_{xy, N}$ for the former and $\delta K^{\rm (SH)}_{xy}$ for the latter. Then, starting from the Kubo formula [Eq.~(\ref{eq:Kxy_func01})] and perform the integral over the position variable, we have for the process shown in Fig.~\ref{fig:triangle01}(a)  
%%%
\begin{eqnarray}
  \delta K^{\rm (SH)}_{xy} (\ui \Omega_m) &=& 
  \frac{1}{2 \pi r_H^2} \sum_{N_1, N_2} \bra N_1 | Q_x | N_2 \ket \bra N_2 | Q_y | N_1 \ket 
  \nonumber \\
  &\times& 
  T \sum_{\omega_{\nu}} t^{(s)}(\ui \omega_{\nu_+}, \ui \omega_\nu) t^{(c)} 
  \nonumber \\
  &\times& \int_q {L}_{N_1, q} (\ui \omega_{\nu_+} ) {L}_{N_2, q} (\ui \omega_\nu ), 
  \label{eq:Kxy_func02}
\end{eqnarray}
%%%
where $\omega_\nu= 2 \pi T \nu$ is a bosonic Matsubara frequency with integer $\nu$, and $\omega_{\nu_+}= \omega_\nu+ \Omega_m$. Here, $r_H= 1/\sqrt{2 |e| H}$ is the magnetic length, $\int_q= \int dq/2\pi$, and the Landau levels $| N \ket$ are defined by
%%%
\begin{equation}
  \widehat{a}_+ \widehat{a}_- | N \ket = N | N \ket,
  \label{eq:Llevel01}
\end{equation}
%%%
where $\widehat{a}_\pm = \frac{r_H}{\sqrt{2}} \left( Q_x \pm \ui Q_y \right)$. In Eq.~(\ref{eq:Kxy_func02}), 
%%%
\begin{equation}
  {L}_{N, q} (\ui \omega_\nu ) = \frac{1}{\alpha_{N,q} + |\omega_\nu|/ \gamma }
  \label{eq:LNq01}
\end{equation}
%%%
is the Matsubara fluctuation propagator of the pair field for the $N$-th Landau level. Here, $\gamma= 8T_{\rm c0}/\pi$, $\alpha_{N, q}= \alpha_N + \xi_0^2 q^2$, and 
%%%
\begin{equation}
  \alpha_N = \frac{T-T_{\rm c0}}{T_{\rm c0}} + h (2N+1), 
  \label{eq:alphaN01}
\end{equation}
%%%
where $h= H/H_{c2}(0)$ with $H_{c2}(0)= 1/(2|e| \xi_0^2)$ being the upper critical field at zero temperature. Using 
%%%
\begin{eqnarray}
  \bra N_1 |Q_x | N_2 \ket &=& \frac{\sqrt{N_1}\delta_{N_1, N_2+1}+  \sqrt{N_2}\delta_{N_1+1,N_2} }{\sqrt{2}r_H}, \\
  \bra N_2 |Q_y | N_1 \ket &=& \frac{\sqrt{N_1} \delta_{N_1, N_2+1}-  \sqrt{N_2}\delta_{N_1+1,N_2}}{-\ui \sqrt{2}r_H}, 
\end{eqnarray}
%%%
which result from $Q_x = (\sqrt{2}r_H)^{-1}(\widehat{a}_++ \widehat{a}_-)$ and $Q_y = (\ui \sqrt{2}r_H)^{-1}(\widehat{a}_+-  \widehat{a}_-)$, we obtain
%%%
\begin{eqnarray}
  \delta K^{\rm (SH)}_{xy} (\ui \Omega_m) &=& 
  \ui T \sum_{N,\omega_{\nu}} \frac{(N+1) t^{(s)}(\ui \omega_{\nu_+}, \ui \omega_\nu) t^{(c)}}{4 \pi r_H^4} \int_q \nonumber \\ 
  &\times& 
  \Big( {L}_{N+1, q} (\ui \omega_{\nu_+} ) {L}_{N, q} (\ui \omega_\nu )
  - (\omega_{\nu_+} \leftrightarrow \omega_\nu) \Big). \nonumber \\
  \label{eq:Kxy_func03}
\end{eqnarray}
%%%

%%%%%%%%%%%%%%%%%%%%%%%%%%%%%%%%%%%%% 
\begin{figure}[t] 
  \begin{center}
    \includegraphics[width=6.0cm]{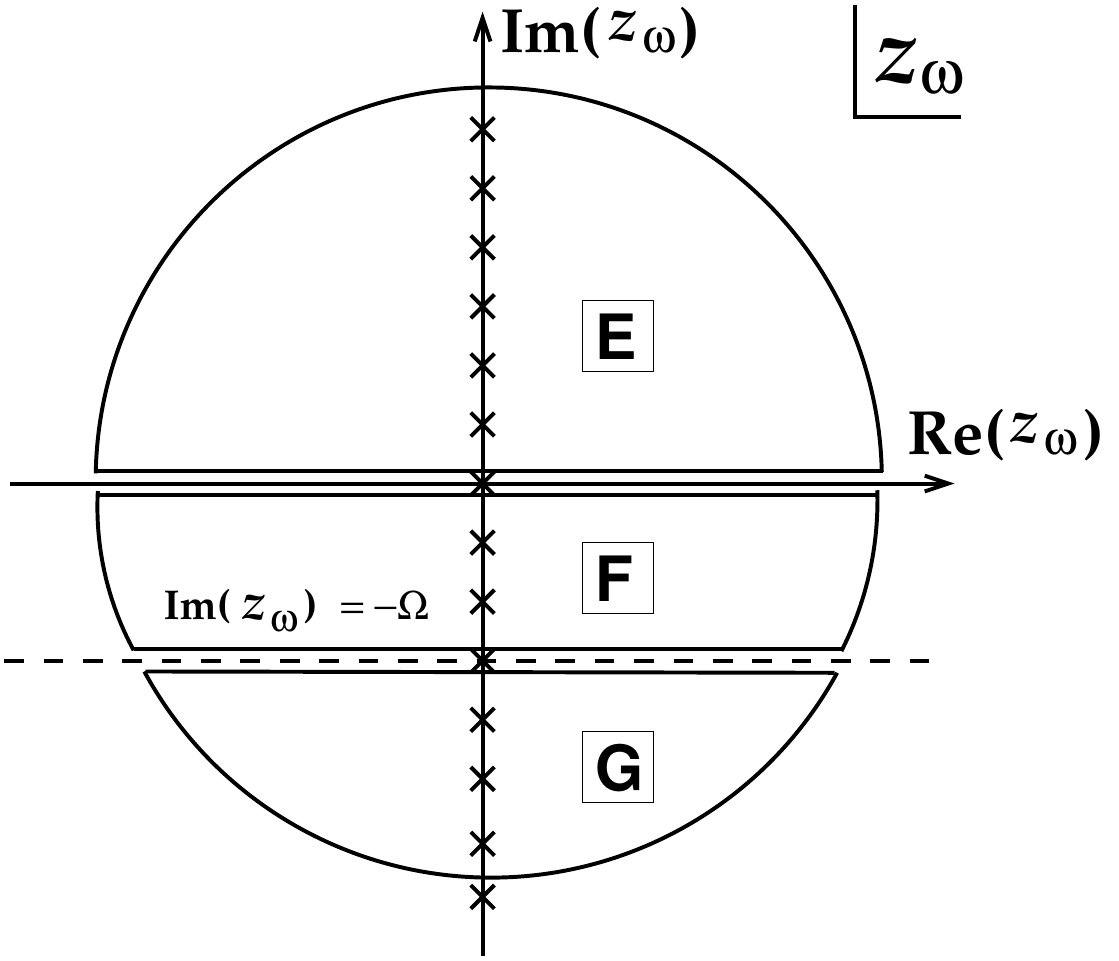}
  \end{center}
  \caption{Contour of integration in the complex $z_\omega$ plane, and three regions \fbox{E}, \fbox{F}, and \fbox{G}.}
  \label{fig:zplane01b}
\end{figure}
%%%%%%%%%%%%%%%%%%%%%%%%%%%%%%%%%%%%%

We next transform the Matsubara sum into the contour integral by using the well-known identity
%%%
\begin{equation}
  T \sum_{\omega_\nu} F(\ui \omega_{\nu}) = \int_C \frac{d z_{\omega}}{4 \pi \ui}
  \coth \left( \frac{z_\omega}{2T}\right) F(z_\omega),  
\end{equation}
%%%
where the contour $C$ is shown in Fig.~\ref{fig:zplane01b}. During the transformation, it is quite important to recall that, as stated at the end of the previous section,  $t^{(s)}$ has a nonzero value only within the \fbox{F} region in Fig.~\ref{fig:zplane01b}. Then, $t^{(s)}$ is analytically continued to $t^{(s)RA}(\omega,\omega)$ in this region \fbox{F}. Substituting this into Eq.~(\ref{eq:Kubo_SH01}), we have 
%%%
\begin{eqnarray}
  \delta \sigma^{\rm (SH)}_{xy} &=& 
   \frac{{t}'^{(s)} t^{(c)} }{8 \pi T r_H^4} \sum_{N} (N+1) \nonumber \\
    & \times& \int_q \int_\omega \frac{\omega}{\sinh^2 \left( \frac{\omega}{2T}\right)} 
   {\rm Im} \left[  {L}^R_{N, q} (\omega) {L}^A_{N+1, q} (\omega) \right],
  \label{eq:sigmaSHxy_01}
\end{eqnarray}
%%% 
where $\int_\omega= \int_{-\infty}^\infty \frac{d \omega}{2 \pi} $.

To proceed further we take the classical limit $\omega/T \ll 1$, which is valid near a superconducting transition at finite temperatures. Then, the integration over $\omega$ can be done as follows: 
%%%
\begin{eqnarray}
  \int_q \int_\omega &&  \frac{\omega}{\sinh^2 \left( \frac{\omega}{2T}\right)} 
      {\rm Im} \left[  {L}^R_{N, q} (\omega) {L}^A_{N+1, q} (\omega) \right] \nonumber \\
      & & \approx 
      2 T^2 \int_q \frac{\alpha_{N+1,q}- \alpha_{N,q}}
          {\alpha_{N,q} \alpha_{N+1,q} (\alpha_{N,q}+ \alpha_{N+1,q})} \nonumber \\
          & & =
          \frac{T^2}{\xi_0 h}
          \left( \frac{1}{\sqrt{\alpha_N}}+ \frac{1}{\sqrt{\alpha_{N+1}}}
          - \frac{2}{\sqrt{\frac{\alpha_N+\alpha_{N+1}}{2}}} \right), 
\end{eqnarray}
%%%
where we used the relation $\alpha_{N+1,q}- \alpha_{N,q}= 2h$. Introducing quantum resistance $R_{\cal Q}= 2 \pi \hbar /4 |e|^2$ (which is written as $\pi/2|e|^2$ in the present unit, $\hbar= 1$), the vortex spin Hall conductivity is finally given by 
%%%
\begin{eqnarray}
  \delta \sigma^{\rm (SH)}_{xy} &=& 
  \frac{\mu_s h}{8 R_{\cal Q} \xi_0 T} \sum_{N} (N+1) \nonumber \\
  && \times \left( \frac{1}{\sqrt{\alpha_N}}+ \frac{1}{\sqrt{\alpha_{N+1}}}
  - \frac{2}{\sqrt{\frac{\alpha_N+\alpha_{N+1}}{2}}} \right), 
  \label{eq:sigmaSHxy_02}
\end{eqnarray}
%%% 
where we used $h= \xi_0^2/r_H^2$.

%%%%%%%%%%%%%%%%%%%%%%%%%%%%%%%%%%%%%%%%%%%%%%%%%%%%
\section{Temperature dependence of vortex spin Hall effect \label{Sec:IV}}
%%%%%%%%%%%%%%%%%%%%%%%%%%%%%%%%%%%%%%%%%%%%%%%%%%%%
In this section, by extending the the Gaussian fluctuation theory [Aslamazov-Larkin theory~\cite{AL68}, Eq.~(\ref{eq:sigmaSHxy_02})] to the non-Gaussian fluctuation theory (self-consistent Hartree approximation~\cite{Ullah91}), we calculate temperature dependence of the vortex spin Hall effect. Here, it is important to note that we are concerned with the vortex fluctuation regime showing a broadening of the resistive transition, where the Landau quantization of the order parameter fixes the length scale of fluctuations {\it transverse} to the applied magnetic field, to the magnetic length $r_H$. Therefore, only the long-wavelength fluctuations {\it along} the magnetic field are left. This leads to the dimensional reduction of the fluctuations, by which the renormalized transition temperature is shifted to a temperature much lower than the mean-field one, and the valid region of fluctuation theory is substantially widened~\cite{Ullah91}.

Before going into details, we briefly discuss the width of the fluctuation region where the self-consistent Hartree approximation, used in the present work, is valid. For a rough estimate of the width of the fluctuation region, we use the Ginzburg criterion~\cite{Chaikin-textbook}: 
%%%
\begin{equation}
  \frac{|T_{\rm c0} - T|}{T_{\rm c0}} \!\alt\! {\rm Gi}
  \label{eq:Ginzburg01}
\end{equation}
%%%
by which the (non-Gaussian) fluctuation region is defined, where ${\rm Gi}$ is the Ginzburg number~\cite{Larkin-textbook}. In a three dimensional superconductor under zero magnetic field, ${\rm Gi}$ is of the order of $10^{-12}$ and hence the fluctuation region is negligibly small~\cite{Larkin-textbook}. By contrast, as mentioned above, fluctuations in the vortex state acquire one-dimensional character. In an effectively one-dimensional superconductor, ${\rm Gi}$ can be of the order of unity~\cite{Larkin-textbook}. Therefore, in the vortex state of a three dimensional superconductor, the fluctuation region spans the whole part of the resistive transition, i.e., from the onset point of the resistivity drop to the accomplished point of zero resistivity. Note, however, that as soon as the vortex pinning effect sets in that establishes the zero resistivity state, the present self-consistent Hartree approximation breaks down. Note also that the self-consistent Hartree approximation deals with the interaction among fluctuations in a non-perturbative way, in that a partial but infinite-order summation of the self-energy for the fluctuation propagator is carried out. 

In the fluctuation region of the vortex state, only the lowest Landau level ($N=0$) becomes critical among other Landau levels, since higher Landau levels ($N>0$) have a gap~\cite{Ikeda89,Tesanovic91}. Because of this fact ($\alpha_{N_1} \gg \alpha_0$ for $N_1 \ge 1$), the non-Gaussian fluctuations substantially renormalize the mass in the lowest Landau level ($\alpha_0$) as 
%%%
\begin{equation}
  \alpha_0 \to \widetilde{\alpha}_0, 
\end{equation}
%%%
whereas the mass in higher Landau levels ($\alpha_{N_1}$) are almost unaffected. Therefore, we can approximate Eq.~(\ref{eq:sigmaSHxy_02}) by 
%%%
\begin{eqnarray}
  \delta \sigma^{\rm (SH)}_{xy} &\approx& 
  \frac{\mu_s h}{8 R_{\cal Q} \xi_0 T} \frac{1}{\sqrt{\widetilde{\alpha}_0}}, 
  \label{eq:sigmaSHxy_03}
\end{eqnarray}
%%%
where the relation $\alpha_{N_1} \gg \widetilde{\alpha}_0$ for $N_1 \ge 1$ has been used. To calculate the renormalized mass $\widetilde{\alpha}_0$ in the lowest Landau level, we adopt the so-called self-consistent Hartree approximation~\cite{Hassing73,Ullah91}, which is known to be a minimal approach that takes into account the fluctuation effects in the superconducting vortex state~\cite{Hassing73,Ullah91}. 

We consider the following Ginzburg-Landau free energy functional: 
%%%
\begin{equation}
  F_{\rm GL} = N(0) \int d^3 r \Big( a |\Psi|^2 + \xi_0^2 |\bmQ \Psi|^2 + \frac{b}{2} |\Psi|^4 \Big),
  \label{eq:FGL01}
\end{equation}
%%%
which are obtained from the BCS Hamiltonian [Eq.~(\ref{eq:H_BCS01})]~\cite{Werthamer-review}. Here, $\Psi$ is the superconducting pair field, $\bmQ$ is the gauge-invariant gradient defined below Eq.~(\ref{eq:gauge-grad01}), and $a= (T- T_{\rm c0})/T_{\rm c0}$, $b= 7 \zeta(3)/(8 \pi^2 T^2)$. To simplify the calculation, we rescale the length and the pair field as $\bmr/\xi_0 \to \bmr$ and $\Psi \sqrt{\frac{N(0)\xi_0^3}{T}} \to \Psi$, so that the free energy functional divided by $T$ now reads 
%%%
\begin{equation}
  \frac{F_{\rm GL}}{T} = \int d^3 r \Big( a |\Psi|^2 + |\bmQ \Psi|^2 + \frac{b'}{2} |\Psi|^4 \Big), 
  \label{eq:FGL02}
\end{equation}
%%%
where the coefficient for the rescaled quartic term, $b'$, is related to the specific heat jump $\Delta C$ under zero field as $b'= {1}/(\Delta C \xi_0^3)$.

%%%%%%%%%%%%%%%%%%%%%%%%%%%%%%%%%%%%% 
\begin{figure}[t] 
  \begin{center}
        \includegraphics[width=9cm]{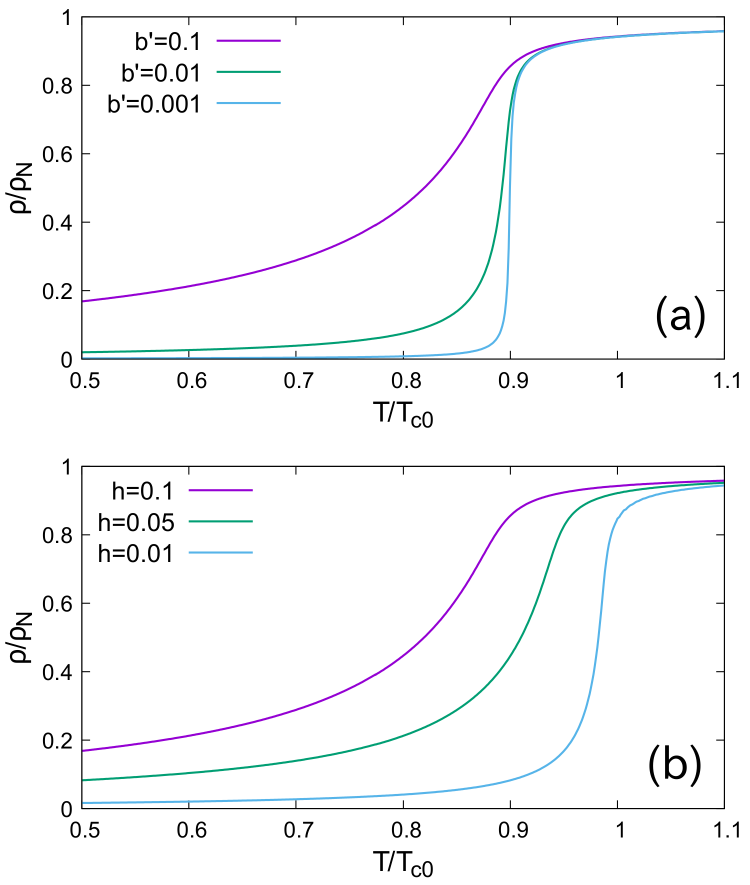}
  \end{center}
  \caption{(a) Temperature dependence of the resistivity $\rho/\rho_N$ for a fixed strength of magnetic field $h=0.1$ with several different values of $b'$. (b) Temperature dependence of $\rho/\rho_N$ for a fixed value $b'=0.1$ [Eq.~(\ref{eq:FGL02})] with several different strengths of magnetic field $h$. In both figures, $\rho_N/(R_{\cal Q} \xi_0)= 0.1$ is used, and the horizontal axis is chosen to be $T/T_{\rm c0}$, where $T_{\rm c0}$ is the transition temperature under zero magnetic field.}   \label{fig:fig_rho-T01}
\end{figure}
%%%%%%%%%%%%%%%%%%%%%%%%%%%%%%%%%%%%%

The orbitals which diagonalize the quadratic terms of the above free energy are the Landau levels. Since only the lowest Landau level ($N=0$) becomes critical in the fluctuation regime as mentioned above~\cite{Ikeda89,Tesanovic91}, and since the lowest Landau level has a macroscopic degeneracy corresponding to the number of vortices $N_v= L_x L_y/2 \pi r_H^2$ with $L_x$ and $L_y $ being the system dimensions in the $x$ and $y$ directions, we expand the pair field into the lowest Landau level orbitals~\cite{Bray74,Ikeda89,Ullah91}: 
%%%
\begin{equation}
  \Psi (\bmr) = \frac{1}{\sqrt{\sqrt{\pi} r_H} L_y L_z}
  \sum_{p,q} \phi_{p,q} u_{p,q} (\bmr),
  \label{eq:LLL01}
\end{equation}
%%%
where we use the Landau gauge $\bmA= Hx \hat{\bf y}$, and $u_{p,q}(\bmr) $ is defined by 
%%%
\begin{equation}
  u_{p,q} (\bmr) = e^{-\frac{1}{2 r_H^2}(x+ p r_H^2)^2+ \ui (p y+ q z) }. 
\end{equation}
%%%
Note that any length here is measured in unit of $\xi_0$. Within the lowest Landau level expansion, the free energy becomes 
%%%
\begin{eqnarray}
  \frac{F_{\rm GL}}{T} &=& \sum_{K} (\alpha_{0}+q^2) \phi^*_{K} \phi_{K} \nonumber \\
  &+& \frac{g}{2} \sum_{K_1, K_2, K_3, K_4} \delta_{K_1+ K_3, K_2+ K_4} \; V_{p_1- p_2} V_{p_3- p_2} \nonumber \\
  && \qquad \qquad \qquad \times \phi^*_{K_1}   \phi_{K_2}   \phi^*_{K_3}   \phi_{K_4},
  \label{eq:FGL_LLL02}
\end{eqnarray}
%%%
where $\alpha_0$ is the $N=0$ component of $\alpha_N$ defined in Eq.~(\ref{eq:alphaN01}), $g= \frac{b'}{\sqrt{2 \pi} r_H L_y L_z}$, $V_{p-p'}= \exp[-\frac{(p-p')^2}{2h}]$, and $K$ stands for a pair $K=(p,q)$.

The self-consistent Hartree approximation~\cite{Hassing73,Ullah91} is a self-consistent one-loop renormalization for the model in Eq.~(\ref{eq:FGL_LLL02}) in the absence of quantum fluctuations. Note that, in the present lowest Landau level approach, vortex fluctuations are naturally taken into account owing to the degeneracy within the lowest Landau level (zeros of the lowest Landau level wave function) that corresponds to the vortex degrees of freedom. After calculating the one-loop self-energy correction for the fluctuation propagator (see Fig.~2(a) of Ref.~\cite{Ikeda89}), we obtain the following self-consistent equation for the mass: 
%%%
\begin{equation}
  \widetilde{\alpha}_0 = \alpha_0 + \frac{h}{2 \pi} \frac{b'}{\sqrt{\widetilde{\alpha}_0}},
  \label{eq:Hartree01}
\end{equation}
%%%
which should be solved for $\widetilde{\alpha}_0$.

Before discussing temperature dependence of the vortex spin Hall conductivity, it is instructive to examine the behavior of diagonal charge conductivity $\sigma_{xx}$, since this quantity is needed later to evaluate the inverse spin Hall voltage. Performing calculations detailed in Appendix~\ref{Sec:App02}, the expression for $\sigma_{xx}$ is given by 
%%%
\begin{eqnarray}
  \sigma_{xx} &\approx& \sigma_N + 
  \frac{\pi}{16 R_{\cal Q} \xi_0 } \frac{1}{\sqrt{\widetilde{\alpha}_0}}, 
  \label{eq:sigmaxx_02}
\end{eqnarray}
%%%
where $\sigma_N$ is the the normal state conductivity. Note that, in the {\it absence} of vortex pinning effects that leads to the vortex glass ordering, not only that the vortex liquid state we are dealing with is a resistive state, but also that the Abrikosov vortex lattice without pinning exhibits the flux flow resistivity in response to the applied electric field~\cite{Rosenstein10}. As emphasized and clarified in Ref.~\cite{Ullah91}, the self-consistent Hartree approximation is applicable to such resistive states, as well as able to describe a smooth crossover from the normal state to the flux flow state.

%%%%%%%%%%%%%%%%%%%%%%%%%%%%%%%%%%%%% 
\begin{figure}[t] 
  \begin{center}
        \includegraphics[width=7cm]{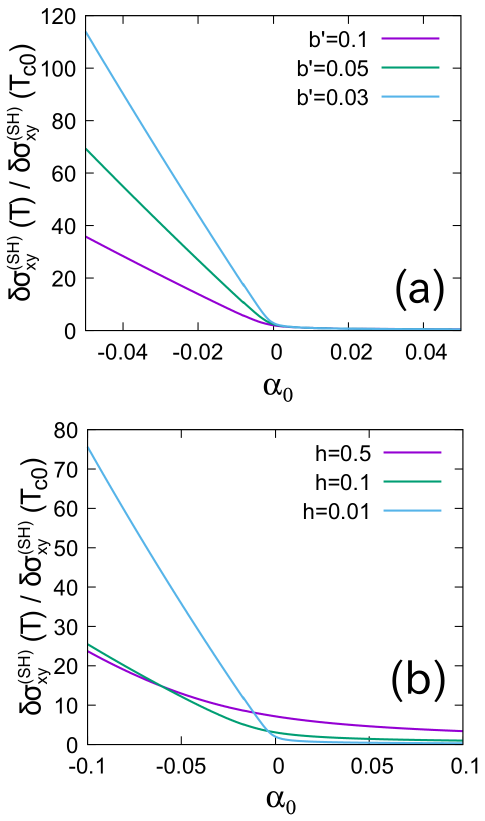}
  \end{center}
  \caption{(a) The vortex spin Hall conductivity $\delta \sigma^{\rm (SH)}_{xy}(T)/\delta \sigma^{\rm (SH)}_{xy}(T_{\rm c0})$ as a function of reduced temperature $\alpha_0= T/T_{\rm c0}-1+ h$ for a fixed strength of the magnetic field $h=0.01$ with several different values of $b'$. (b) $\delta \sigma^{\rm (SH)}_{xy}(T)/\delta \sigma^{\rm (SH)}_{xy}(T_{\rm c0})$ for a fixed $b'=0.1$ value with several different strengths of the magnetic field $h$. 
  }  
  \label{fig:fig_sigmaSxy01}
\end{figure}
%%%%%%%%%%%%%%%%%%%%%%%%%%%%%%%%%%%%%

%%%%%%%%%%%%%%%%%%%%%%%%%%%%%%%%%%%%% 
\begin{figure*}[t] 
  \begin{center}
        \includegraphics[width=18.0cm]{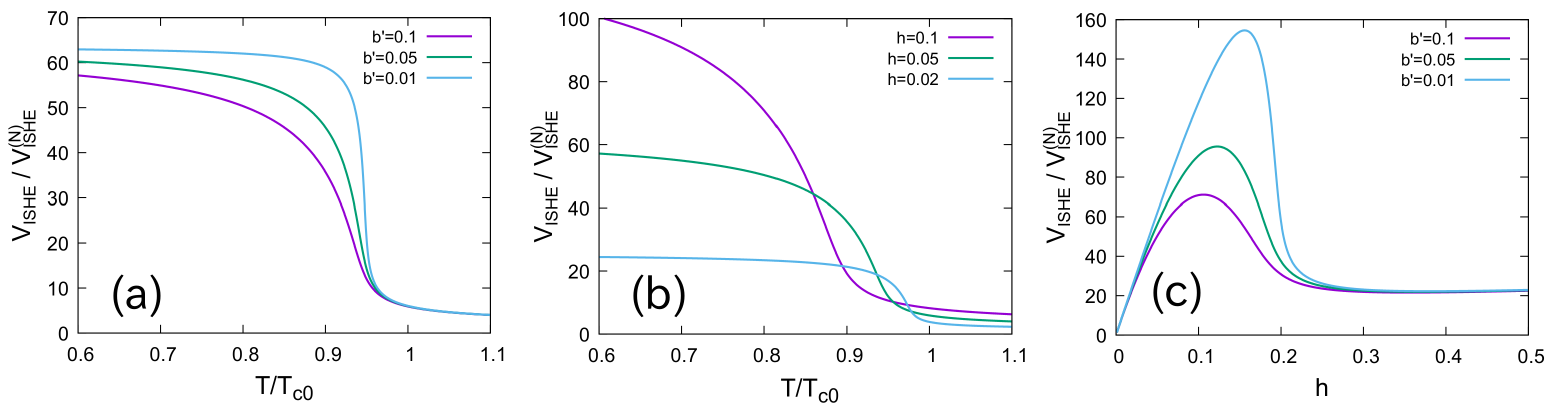}
  \end{center}
  \caption{(a) Temperature dependence of the inverse spin Hall voltage $V_{\rm ISHE}$, normalized by the normal state value $V_{\rm ISHE}^{(N)}$, for a fixed strength of the magnetic field $h=0.05$ with several different values of $b'$. Here, $\rho_N/(R_{\cal Q} \xi_0)=0.1$, $\mu_s/T_{\rm c0}=2.0$, and $\theta^{\rm (SH)}_N= 0.001$ are used, where $\theta^{\rm (SH)}_N$ is the spin Hall angle in the normal state. (b) Temperature dependence of $V_{\rm ISHE}$ for a fixed value $b'=0.1$ with several different strengths of the magnetic field $h$. (c) Magnetic field dependence of $V_{\rm ISHE}$ at $T/T_{\rm c0}=0.8$ for several different choices of $b'$.
    Note that the vortex pinning effects are disregarded in all plots. 
  } 
  \label{fig:fig_VISHE-h}
\end{figure*}
%%%%%%%%%%%%%%%%%%%%%%%%%%%%%%%%%%%%%

Figure~\ref{fig:fig_rho-T01} shows the resistivity $\rho= 1/\sigma_{xx}$ calculated using Eq.~(\ref{eq:sigmaxx_02}) as a function of temperature, where the resistivity is normalized by the normal state value $\rho_N= 1/\sigma_N$. In viewing the figure, it is important to recall that the mean-field transition temperature under magnetic field is given by $T^{\rm MFA}_{c}(h)/T_{\rm c0}= 1-h$, and that the lower temperature side below $T^{\rm MFA}_{c}(h)$ is still a fluctuation region due to the dimensional reduction of fluctuations as emphasized at the beginning of this section. This fluctuation region, i.e., the vortex liquid region, can be well discribed by the self-consistent Hartree approximation~\cite{Ullah91}.

In Fig.~\ref{fig:fig_rho-T01}(a), temperature dependence of $\rho/\rho_N$ is plotted for a fixed strength of magnetic field $h$ with several different choices of the coefficient $b'$ in Eq.~(\ref{eq:FGL02}). Since the $b'$ value is known to measure the magnitude of the fluctuations~\cite{Chaikin-textbook}, with the increase of $b'$, the fluctuation region is widened. In Fig.~\ref{fig:fig_rho-T01}(b), the resistivity is plotted for a fixed value of $b'$ [Eq.~(\ref{eq:FGL02})] with several different strengths of the magnetic field $h$. From the figure we see that, with increasing the magnetic field, the onset of superconducting fluctuation is shifted to lower temperatures, as well as the width of the fluctuation region is enlarged. Note that such behaviors are well known not only in cuprate superconductors~\cite{Ikeda91} but also in conventional superconductors~\cite{Pourret09} (see Fig. 2 therein). The present calculation qualitatively captures these features.

Now we discuss the vortex spin Hall conductivity $\delta \sigma^{\rm (SH)}_{xy}$. Figure~\ref{fig:fig_sigmaSxy01} shows the vortex spin Hall conductivity calculated using Eqs.~(\ref{eq:sigmaSHxy_03}) and (\ref{eq:Hartree01}) as a function of reduced temperature $\alpha_0= T/T_{\rm c0}-1+ h$. In Fig.~\ref{fig:fig_sigmaSxy01}(a), the $b'$ value is varied as $b'= 0.1, 0.05, 0.03$ while the strength of the magnetic field $h=0.01$ is fixed. From Fig.~\ref{fig:fig_sigmaSxy01}(a) we see that the magnitude of $b'$ does not affect the fluctuation region of the vortex spin Hall conductivity so much, but it controls the growth of $\delta \sigma^{\rm (SH)}_{xy}$ below the mean-field transition temperature ($\alpha_0= 0$). In Fig.~\ref{fig:fig_sigmaSxy01}(b), the strength of the magnetic field is varied as $h= 0.5, 0.1, 0.01$ while the value $b'=0.1$ is kept constant. From the figure we find that the strength of the magnetic field enlarges the fluctuation region as well as the magnitude of the vortex spin Hall conductivity.

Next, we discuss temperature dependence of the inverse spin Hall voltage, i.e., electric voltage produced by the inverse spin Hall effect, which can be directly compared with experiments. The linear response equation for the charge current $j^{(c)}_x$ in the presence of electric field $E_x$ and spin voltage gradient $\nabla_y \mu_s$ is given by~\cite{Takahashi08} 
%%%
\begin{equation}
  j^{(c)}_x = \sigma_{xx} E_x + \sigma^{\rm (SH)}_{xy} (- \nabla_y \mu_s), 
\end{equation}
%%%
where $E_x$ is the $x$-component of the electric field. Under the open circuit condition $j^{(c)}_x=0$, the electric voltage produced by the inverse spin Hall effect is given by
%%%
\begin{equation}
  V_{\rm ISHE} = \frac{w \sigma^{\rm (SH)}_{xy}}{\sigma_{xx} \sigma^{\rm (s)}} j^{(s)}_y, 
  \label{eq:VISHE01}
\end{equation} 
%%%
where $w$ is the distance of the two contacts to detect the Hall voltage, and $j^{(s)}_y= \sigma^{\rm (s)} (-\nabla_y \mu_s)$ is the injected spin current. Note that there is no divergent contribution to the diagonal spin conductivity $\sigma^{\rm (s)}$ from superconducting fluctuations, as the situation is similar to the absence of divergent contribution to the thermal conductivity from superconducting fluctuations~\cite{Vishveshwara01,Niven02}. Therefore, we can safely set $\sigma^{\rm (s)}= \sigma_N$ in the vortex fluctuation region. Keeping this note in mind and substituting Eq.~(\ref{eq:sigmaSH01}) into Eq.~(\ref{eq:VISHE01}), the inverse spin Hall voltage can be rewritten as
%%%
\begin{equation}
  V_{\rm ISHE} = \frac{w}{\sigma_{xx}} \left( \theta^{\rm (SH)}_N+ \frac{\delta \sigma^{\rm (SH)}_{xy}}{\sigma_N} \right)j^{(s)}_y, 
  \label{eq:VISHE02}
\end{equation} 
%%%
where the first term on the right-hand side arises from the normal-state spin Hall conductivity, 
%%%
\begin{equation}
  \sigma^{\rm (SH)}_{xy,N} =  \theta^{\rm (SH)}_{N} \sigma_N, 
\end{equation}
%%%
where $\theta^{\rm (SH)}_{N}$ is the spin Hall angle in the normal state. Eq.~(\ref{eq:VISHE02}) means that, in the absence of the spin-orbit interaction, there is no normal state contribution to the spin-Hall conductivity, such that the vortex spin Hall effect is dominant as in the case of vortex Nernst/Ettingshausen effect. In the presence of the spin-orbit interaction, however, the ratio of the vortex contribution to the normal state contribution depends on the magnitude of the spin Hall angle in the normal state.  

Figure~\ref{fig:fig_VISHE-h}(a) shows temperature dependence of the inverse spin Hall voltage $V_{\rm ISHE}$ calculated from Eq.~(\ref{eq:VISHE02}) for a fixed strength of the magnetic field $h=0.05$, whereas Fig.~\ref{fig:fig_VISHE-h}(b) shows the same quantity for a fixed value of $b'=0.1$. From these two figures we find that the general tendency of $V_{\rm ISHE}$ is essentially the same as that in Figs.~\ref{fig:fig_rho-T01} and \ref{fig:fig_sigmaSxy01}, namely, an increase of the magnetic field $h$ as well as that of the coefficient $b'$ enlarges the fluctuation region of $V_{\rm ISHE}$, which makes an experimental detection of the signal more tractable. Besides, in Fig.~\ref{fig:fig_VISHE-h}(c), we show magnetic field dependence of $V_{\rm ISHE}$ at a fixed temperature $T/T_{\rm c0}= 0.8$. Recalling that the upper critical field in the mean-field approximation is given by $h^{\rm MFA}_{c2}= 0.2$ at $T/T_{\rm c0}= 0.8$, a magnetic-field window below $h^{\rm MFA}_{c2}$ still belongs to a fluctuation region due to the dimensional reduction of fluctuations as discussed at the beginning of this section. Therefore, in superconductors without vortex pinning effects, the peak seen in Fig.~\ref{fig:fig_VISHE-h}(c) is within the valid region of the present self-consistent Hartree approximation. Note however that, when comparing the theoretical result with experiments, care must be taken that vortex pinning effects that exist in a real material may reduce the signal at lower fields, thus narrowing the width of the peak.  

\begin{figure}[t] 
  \begin{center}
        \includegraphics[width=8cm]{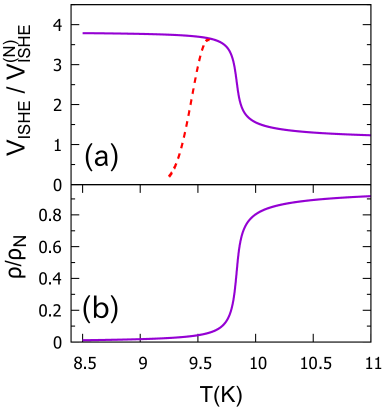} 
  \end{center}
  \caption{(a) Temperature dependence of the inverse spin Hall voltage $V_{\rm ISHE}$, normalized by the normal state value $V_{\rm ISHE}^{(N)}$, for a fixed strength of the magnetic field $h=0.015$. Here, $T_{\rm c0}= 10.0$~K, $b'=0.02$, $\rho_N/(R_{\cal Q} \xi_0)=0.155$, $\mu_s/T_{\rm c0}=2.0$, and $\theta^{\rm (SH)}_N= 0.005$ are used~\cite{Morota11,Jeon18b,Rogdakis19}, where $\theta^{\rm (SH)}_N$ is the spin Hall angle in the normal state. The solid line is a result calculated using Eq.~(\ref{eq:VISHE02}), whereas the dotted line is an {\it expected} behavior once taking account of the vortex pinning effects (see the main text). (b) Temperature dependence of the resistivity with the same parameter as (a). 
  } 
  \label{fig:fig_VISHE01}
\end{figure}

Finally, in Fig.~\ref{fig:fig_VISHE01}, we try to interpret the experimental result of Ref.~\cite{Umeda18} in terms of the vortex spin Hall effect. Figure~\ref{fig:fig_VISHE01}(a) shows temperature dependence of the inverse spin Hall voltage $V_{\rm ISHE}$ calculated from Eq.~(\ref{eq:VISHE02}), which includes the spin Hall effect in the normal state of NbN~\cite{Morota11,Jeon18b,Rogdakis19}. Here, the parameters are chosen in such a way that the result is comparable with Ref.~\cite{Umeda18}. (See the resistivity calculation in Fig.~\ref{fig:fig_VISHE01}(b), which should be compared with Fig.~3(a) of Ref.~\cite{Umeda18}.) From the result for $V_{\rm ISHE}$ [solid line in Fig.~\ref{fig:fig_VISHE01}(a)] we find that, with decreasing temperature, there is a steep growth of $V_{\rm ISHE}$ accompanied by a saturation at lower temperatures, whose features are consistent with Ref.~\cite{Umeda18}. In the figure, we also draw an expected behavior when we take into account the vortex pinning effects, which forms a peak around the superconducting transition [dotted line in Fig.~\ref{fig:fig_VISHE01}(a)]. 

This drawing (dotted line) is based on the following arguments. First, the (inverse) vortex spin Hall effect under discussion is a process where the spin diffusion concomitant with the vortex motion produces a time-dependent phase variation of the pair-field, thus generating the transverse voltage in accordance with the Josephson equation~\cite{Josephson62,Anderson66}. Therefore, once the vortex pinning effect sets in, the vortex motion and the phase variation start to freeze, leading to a decrease in the transverse voltage. Second, as already pointed out in Sec.~\ref{Sec:I}, there is a strong similarity between the vortex spin Hall effect and the vortex Nernst/Ettingshausen effects. In the case of the vortex Nernst effect, the voltage signal is known to form a peak as a function of temperature~\cite{Pourret09}. Therefore, considering the similarity to the vortex Nernst effect, the inverse spin Hall voltage due to the vortex spin Hall effect is expected to show a peak structure as drawn by the dotted line. 

As emphasized in Sec.~\ref{Sec:I}, the calculated temperature dependence of the inverse spin Hall voltage [Fig.~\ref{fig:fig_VISHE01}(a)] may provide a possible explanation for a prominent peak found in the spin Seebeck effect in a NbN/Y$_3$Fe$_5$O$_{12}$ system~\cite{Umeda18}.

%%%%%%%%%%%%%%%%%%%%%%%%%%%%%%%%%%%%% 
%%%%%%%%%%%%%%%%%%%%%%%%%%%%%%%%%%%%%

%%%%%%%%%%%%%%%%%%%%%%%%%%%%%%%%%%%%%%%%%%%%%%%%%%%%
\section{Discussion and conclusion \label{Sec:V}}
%%%%%%%%%%%%%%%%%%%%%%%%%%%%%%%%%%%%%%%%%%%%%%%%%%%%
In this paper, we have theoretically examined the spin Hall effect driven by the motion of superconducting vortices, which we term the vortex spin Hall effect. Focusing on the influence of superconducting fluctuations that are known to play a crucial role in the resistive transition of type-II superconductors, we have clarified the following two points. First, we have shown through a diagrammatic calculation that there is a strong similarity between the vortex spin Hall effect under discussion and the well-known vortex Nernst/Ettingshausen effect~\cite{Wang06,Pourret06}. Second, we have demonstrated that the spin Seebeck effect in a superconductor/ferromagnetic insulator system can exhibit a prominent peak as a function of temperature~\cite{Umeda18}, not due to the coherence peak effect~\cite{Inoue17}, but due solely to the (inverse) vortex spin Hall effect. The same conclusion applies to the inverse spin Hall voltage for a spin pumping in a superconductor/ferromagnetic insulator system.

  The vortex spin Hall effect which we have examined here offers a new means of spin-to-charge current interconversion~\cite{Otani17,Han18} in superconductors. As is seen from Figs.~\ref{fig:fig_VISHE-h} and \ref{fig:fig_VISHE01}, this effect is most conspicuous in the fluctuation region of the superconducting vortex state where the vortex pinning effects do not set in. Since a superconductor with strong fluctuations exhibits a wider fluctuation region, it is tempting to carry out a spin pumping/Seebeck experiment using cuprate superconductors or iron-based superconductors.

We emphasize that the vortex spin Hall effect does not require the spin-orbit scattering, but requires a non-zero spin accumulation inside the vortex core. Therefore, the signal is not proportional to the strength of the spin-orbit interaction in the system, but proportional to the magnitude of the spin accumulation. In this respect, in a superconductor with no spin-orbit interactions, the vortex spin Hall effect can be easily identified because there is no normal state contribution. 

To summarize, we have developed a microscopic theory of the vortex spin Hall effect, i.e., the spin Hall effect driven by the motion of superconducting vortices. We have pointed out a strong similarity between the vortex spin Hall effect and the well-known vortex Nernst/Ettingshausen effect, and shown that the spin Seebeck effect in a ferromagnetic insulator/superconductor system can exhibit a prominent peak as a function of temperature~\cite{Umeda18}. Since the predicted signal is most visible in a superconductor with strong fluctuations, we hope that a spin pumping/Seebeck experiment using cuprate or iron-based superconductors is performed in future.

%%%
\acknowledgments 
%%%
This work was financially supported by JSPS KAKENHI Grant Number 19K05253.

\appendix

%%%%%%%%%%%%%%%%%%%%%%%%%%%%%%%%%%%%%%%%%%%%%%%%%%%%
\section{Derivation of the spin current vertex \label{Sec:App01}}
%%%%%%%%%%%%%%%%%%%%%%%%%%%%%%%%%%%%%%%%%%%%%%%%%%%%

In this Appendix, we sketch the derivation of the spin current vertex [Eq.~(\ref{eq:Js_02})]. We first consider the case $\omega_1 > 0$, $\omega_2 <0$ [Fig.~\ref{fig:zeps01}(a)]. The momentum integral for the product of three Green's functions yields $- 4 \pi N(0) \tau^2 D \bmQ$ as in the case of charge current vertex. Then, after considering a contribution from the two Cooperons, we obtain 
%%%
\begin{equation}
  {\bf J}^{(s)} = {-4 \pi |e|}N(0) D \bmQ ( {\cal S}_A + {\cal S}_D ),
  \label{eq:Js_A01}
\end{equation}
%%%
where 
%%%
\begin{eqnarray}
  {\cal S}_A &=& \sum_{-\omega_2 < \eps_n < \infty, \sigma} 
  \frac{\sigma T}{(|2 \eps_n + \omega_1|-\ui \sigma \mu_s)
    (|2 \eps_n + \omega_2|-\ui \sigma \mu_s)} \nonumber
\end{eqnarray}
%%%
comes from region \fbox{A} in Fig.~\ref{fig:zeps01}(a), whereas 
%%%
\begin{eqnarray}
       {\cal S}_D &=& \sum_{-\infty < \eps_n < -\omega_1, \sigma} 
       \frac{\sigma T}{(|2 \eps_n + \omega_1|+\ui \sigma \mu_s)
         (|2 \eps_n + \omega_2|+\ui \sigma \mu_s)}  \nonumber 
\end{eqnarray}
%%%
comes from region \fbox{D}. After shifting the Matsubara frequency $\eps_n \to \eps_n - \omega_2$, ${\cal S}_A$ is transformed to 
%%%
\begin{eqnarray}
  {\cal S}_A &=& \sum_{0 < \eps_n < \infty, \sigma} 
  \frac{\sigma T}{(2 \eps_n - \omega_2+ \omega_{12} -\ui \sigma \mu_s)
    (2 \eps_n - \omega_2-\ui \sigma \mu_s)} \nonumber \\
  &=& \frac{T}{(4 \pi T)^2} \sum_\sigma 
  \sigma \varPsi^{(1)} \left(\frac{1}{2}+ \frac{- \omega_2 -\ui \sigma \mu_s}{4 \pi T} \right), 
\end{eqnarray}
%%%
where $\omega_{12}= \omega_1- \omega_2$, which becomes zero upon analytic continuation $\omega_{1,2} \to - \ui \omega \pm 0_+$ and hence can be safely set to zero in the argument of tri-gamma function. Similarly, after reversing the sign of the Matsubara frequency $\eps_n \to -\eps_n $, ${\cal S}_D$ is transformed to 
%%%
\begin{eqnarray}
  {\cal S}_D &=& \sum_{\omega_1 < \eps_n < \infty, \sigma} 
  \frac{\sigma T}{(2 \eps_n - \omega_2 + \ui \sigma \mu_s) (2 \eps_n - \omega_1 + \ui \sigma \mu_s)}, \nonumber  
\end{eqnarray}
%%%
which corresponds to the expression of ${\cal S}_A$ under the exchange $\sigma \mu_s \leftrightarrow -\sigma \mu_s$ and $\omega_1 \leftrightarrow -\omega_2$. Therefore, ${\cal S}_D$ is obtained as 
%%%
\begin{eqnarray}
  {\cal S}_D &=& \frac{T}{(4 \pi T)^2}\sum_\sigma
  \sigma \varPsi^{(1)} \left(\frac{1}{2}+ \frac{\omega_1 +\ui \sigma \mu_s}{4 \pi T} \right). 
\end{eqnarray}
%%%
Then, after the analytic continuation $\omega_{1,2} \to - \ui \omega \pm 0_+$, the sum ${\cal S}_A+ {\cal S}_D$ is calculated to be 
%%%
\begin{eqnarray}
  {\cal S}_A + {\cal S}_D &=& 
  \frac{T}{(4 \pi T)^2} \sum_\sigma \sigma \Bigg[
    \varPsi^{(1)} \left( \frac{1}{2}+ \frac{\ui \omega - \ui \sigma \mu_s }{4 \pi T} \right) \nonumber \\
    && +
    \varPsi^{(1)} \left( \frac{1}{2}+ \frac{-\ui \omega + \ui \sigma \mu_s }{4 \pi T} \right) \Bigg] \nonumber \\
  &=& \frac{4T \mu_s \omega}{(4 \pi T)^4} \varPsi^{(3)} \left( \frac{1}{2} \right), 
\end{eqnarray}
%%%
where $\varPsi^{(3)}$ is the penta-gamma function, and we expanded the result to linear order in $\omega$ as well as in $\mu_s$.

In the case $\omega_1, \omega_2 >0$ [Fig.~\ref{fig:zeps01}(b)], the contributions from region \fbox{A} and \fbox{D} cancel as long as we are concerned with the spin current vertex proportional to the frequency $\omega_\nu$. Using $\varPsi^{(3)}(1/2)= \pi^4$, we obtain Eq.~(\ref{eq:Js_02}) in the main text.  

%%%%%%%%%%%%%%%%%%%%%%%%%%%%%%%%%%%%%%%%%%%%%%%%%%%%
\section{Calculation of the fluctuation conductivity \label{Sec:App02}}
%%%%%%%%%%%%%%%%%%%%%%%%%%%%%%%%%%%%%%%%%%%%%%%%%%%%
Here, we review the calculation of the fluctuation charge conductivity~\cite{AL68}: 
%%%
\begin{equation}
  \sigma_{xx} = \frac{1}{\ui \Omega} \left. \left( K^{R}_{xx} (\Omega) - K^{R}_{xx} (0) \right) \right|_{\Omega \to 0}, 
  \label{eq:sigma_xx01}
\end{equation}
%%%
where the retarded current correlation function $K^{R}_{xx} (\Omega)$ is obtained from the Matsubara function, 
%%%
\begin{eqnarray}
  K_{xx} (\ui \Omega_{m}) 
  &=&
  -\frac{1}{V}  \int d^3 r \int d^3 r'  
  \int_0^{\beta} d u e^{\ui \Omega_{m} u} \nonumber \\
  && \times \bra T_u \; {\mathcal J}^{(c)}_x (\bmr, u) {\mathcal J}^{(c)}_x (\bmr', 0) \ket
  \label{eq:Kxx_func01}
\end{eqnarray}
%%%
by an analytic continuation $\ui \Omega_m \to \Omega + \ui 0_+$. 

As in the case of the spin Hall conductivity [Eq.~(\ref{eq:sigmaSH01})], the diagonal conductivity can be divided into the normal state contribution $\sigma_{xx,N}$ and the superconducting fluctuation contribution $\delta \sigma_{xx}$ as
%%%
\begin{equation}
  \sigma_{xx} = \sigma_{N}+ \delta \sigma_{xx},
  \label{eq:sigma01}
\end{equation}
%%%
where the former (latter) is obtained from the correlation function $K_{N}$ ($\delta K_{xx}$). Writing down the process shown in Fig.~\ref{fig:triangle01}(a) with the current vertex given in Sec.~\ref{Sec:II}, we have 
%%%
\begin{eqnarray}
  \delta K_{xx} (\ui \Omega_m) &=& 
  \frac{(t^{(c)})^2}{2 \pi r_H^2} T \sum_{\omega_{\nu}} 
  \sum_{N_1, N_2} \bra N_1 | Q_x | N_2 \ket \bra N_2 | Q_x | N_1 \ket \nonumber \\
  && \times \int_q {L}_{N_1, q} (\ui \omega_{\nu_+} ) {L}_{N_2, q} (\ui \omega_\nu ) \nonumber \\
  &=& \frac{(t^{(c)})^2}{4 \pi r_H^4} \sum_{N} (N+1)
  T \sum_{\omega_{\nu}}\nonumber \\
  &\times&  \int_q  \Big( {L}_{N+1, q} (\ui \omega_{\nu_+} ) {L}_{N, q} (\ui \omega_\nu )
  + (\omega_{\nu_+} \leftrightarrow \omega_\nu) \Big), \nonumber \\
  \label{eq:Kxx_func02}
\end{eqnarray}
%%%
where $\omega_\nu= 2 \pi T \nu$ is a bosonic Matsubara frequency with integer $\nu$, and $\omega_{\nu_+}= \omega_\nu+ \Omega_m$. After performing the Matsubara sum as well as the analytic continuation, we obtain 
%%%
\begin{eqnarray}
  \delta \sigma_{xx} &=&    \frac{(t^{(c)} )^2}{4 \pi r_H^4 T} 
  \sum_{N} (N+1) \nonumber \\
  &\times& 
  \int_q \int_\omega \frac{1}{\sinh^2 \left( \frac{\omega}{2T} \right)} 
  {\rm Im} L_{N,q}^{R}(\omega)   {\rm Im} L_{N+1,q}^{A}(\omega). 
\end{eqnarray}
%%%
Taking the classical limit $\omega/T \ll 1$ and then performing the integral over $\omega$ and $q$, we obtain 
%%%
\begin{eqnarray}
  \delta \sigma_{xx} &=& 
  \frac{\pi}{16 R_{\cal Q} \xi_0 } \sum_{N} (N+1) \nonumber \\
  && \times \left( \frac{1}{\sqrt{\alpha_N}}+ \frac{1}{\sqrt{\alpha_{N+1}}}
  - \frac{2}{\sqrt{\frac{\alpha_N+\alpha_{N+1}}{2}}} \right). 
  \label{eq:sigmaxx_01}
\end{eqnarray}
%%%
Using the lowest Landau level approximation ($\alpha_{N_1} \gg \alpha_0$ for $N_1 \ge 1$), and combining Eq.~(\ref{eq:sigmaxx_01}) with Eq.~(\ref{eq:sigma01}), we obtain Eq.~(\ref{eq:sigmaxx_02}) in the main text.

%%%%%%%%%%%%%%%%%%%%%%%%%%%%%%%%%%%%%%%%%%%%%%%%%%%%%%%%%%%%%%%%%%%%%%%%%%%%%%%%%%%%%%%
% Create the reference section using BibTeX:
%\bibliography{basename of .bib file}

%\clearpage

%\subsection*{Figure Captions} 

\end{document}